\documentclass[12pt,preprint]{aastex}

\def\lsim{\hbox{\raise.35ex\rlap{$<$}\lower.6ex\hbox{$\sim$}\ }}

\newcommand{\rosat}{{\it ROSAT}}

\newcommand{\asca}{{\it ASCA}}

\newcommand{\myarcmin}{^\prime\mskip-5mu}

\slugcomment{For submission to {\it The Astrophysical Journal}}

\received{2003 January 3}
\begin{document}

\title{An X-ray study of the supernova remnant G18.95$-$1.1}
\author{I. M.~Harrus}
\affil{NASA/USRA Goddard Space Flight Center, Greenbelt MD 20771}
\email{imh@lheapop.gsfc.nasa.gov}

\author{P. O. Slane}
\affil{Harvard-Smithsonian Center for Astrophysics, Cambridge MA 02138}

\author{J. P. Hughes}
\affil{Rutgers University, Piscataway NJ 08854}

\and
\author{P. P. Plucinsky}
\affil{Harvard-Smithsonian Center for Astrophysics, Cambridge MA 02138}

\begin{abstract}
We present an analysis of data from both the {\it R\"{o}ntgen Satellite}
(\rosat), and the {\it Advanced
Satellite for Cosmology and Astrophysics} (\asca) 
of the supernova remnant (SNR) G18.95$-$1.1.  
We find that the X-ray emission from G18.95$-$1.1 is predominantly thermal,  heavily 
absorbed with a column density around 10$^{22}$~atoms~cm$^{-2}$ and can be  best described by an NEI (nonequilibrium ionization) model with a temperature around 0.9~keV, and an ionization timescale of  
1.1$\times$10$^{10}$~cm$^{-3}$~s$^{-1}$.  We find only marginal evidence for non-solar abundances. Comparisons between 21 cm HI absorption data and derived parameters from our spectral analysis strongly suggest a  relatively near-by remnant (a distance of about 2~kpc).

Above 4~keV, we identify a small region of emission located at the tip of the central, flat spectrum bar-like feature in the radio image. We examine two possibilities for this emission region: a temperature variation within the remnant or a pulsar wind nebula (PWN). The current data do not allow us to distinguish between these possible explanations. In the scenario where this high-energy emission region corresponds to a PWN, our analysis suggests a rotational loss rate for the unseen pulsar of  about  7$\times$10$^{\rm 35}$ erg~s$^{\rm -1}$  and a ratio $L_{r}$/$L_{x}$ about  3.6  for the entire PWN, slightly above the maximum ratio (3.4 for Vela) measured in known PWN. 
\end{abstract}


\section{Introduction}

G18.95$-$1.1 was discovered as a non-thermal extended radio source during 
a continuum survey of the Galactic plane \citep{rei84}. 
The source was first suggested to be energized by a binary system containing a compact
object \citep{fur85}.
This classification was based on the unusual radio morphology of 
G18.95$-$1.1, 
consisting of various arcs pointing toward a central radio peak.  
Follow-up radio observations \citep{ode86} were done soon afterwards to 
confirm this identification. Based on 
a detailed analysis of the variation of the spectral index  
across G18.95$-$ 1.1 and on a re-analysis of the morphological 
characteristics used by \citet{fur85},
\citet{ode86} concluded that G18.95$-$1.1 was more likely a low surface 
brightness Crab-like supernova remnant (SNR) powered  
by a hidden central compact object, although the binary interpretation could not
be completely ruled-out. He suggested further 
 high-resolution low-frequency observations at 327~MHz to 
resolve the ``bar-like'' structure at the center of the source. 
Such observations were carried out by \citet{pat88}, and their analysis 
yielded the detection of an almost complete shell surrounding a compact,
central object. This result established that G18.95$-$1.1
 was indeed a supernova remnant, but the origin of the 
energy sustaining the central emission from G18.95$-$1.1 remained unknown.  
\citet{ode86} suggested that the SNR was powered by a central pulsar. However, \citet{fur89} argued that  G18.95$-$1.1 could be a binary system consisting of the neutron star 
created in the supernova explosion and a low mass companion unseen because of the known large optical absorption. To solve the mystery of the energy source of the radio emission and search for a possible central object, 
a pointed \rosat\ PSPC observation (12~ks) of the SNR was carried out
by \citet{fur97}. 
They found that  the remnant could be best described by a thermal emission model 
\citep{ray77} at a temperature of about 0.95~keV and a hydrogen column density of 3.4$\pm$1.5$\times$10$^{21}$~atoms~cm$^{-2}$.  A non-thermal model (power-law) yielded a photon index of $\alpha$~=~$-$8.9 which was therefore dismissed as an ``unrealistic" model. 
\setcounter{footnote}{0}
In addition to the X-ray data, \citet{fur97} obtained a 
detailed radio map (at an angular resolution of 
69$^{\prime\prime}$) using the Effelsberg 100-m telescope at 10.55 GHz where 
Faraday depolarization effects are small.
The image extracted from these data is shown in 
Figure~1\footnote{The data were kindly provided by Professor F\"{u}rst at the 
Max-Planck-Institute f\"ur Radioastronomie.}. 
It shows  a distinct central bar  surrounded
by a faint shell of diffuse emission. 
About 80\% of the radio emission 
is in the central diffuse component and has a total flux  of
about 20.4$\pm$0.2~Jy at 10.55~GHz.
This result is consistent with 
existing radio data at 1.4~GHz and 4.75~GHz for a spectral index ($S_\nu \propto \nu^\alpha$)  
$\alpha$~=~$-$0.14$\pm$0.03 for the diffuse component and 
$\alpha$~=~$-$0.22$\pm$0.07 for the central bar. A prominent arc on the western part of the remnant (see Figure~1) has a steeper 
spectral index of $\alpha$~=~$-$0.36$\pm$0.04.  
The linear polarization across the remnant is about 6\% at 10.55~GHz, a little 
more than twice that at 4.75~GHz. 
The polarization intensity is highly concentrated on small scales: when the
 diffuse contribution is subtracted, the polarization  
increases to about 40\% in the central bar and in the arc. 

As is often the case in SNR studies, 
the distance to G18.95$-$1.1 is not well known. 
There exist 21~cm HI absorption data \citep{bra84} 
which were re-analyzed by \citet{fur89}. 
Based on a radial velocity of 18~km~s$^{\rm -1}$ associated with G18.95$-$1.1, they 
deduced a distance of either 2~kpc (within the Sagittarius arm) 
or 15~kpc (on the far side of the Galaxy). 
Because of the large luminosity and 
velocity expansion implied by this latter 
distance, they considered the closer distance
as the most probable and derived all physical quantities using a
2~kpc measure to the SNR. We 
will derive all our results normalized to this 
distance unless otherwise specified and we comment at the end of the paper on the possibility 
that the remnant is as far away as 15~kpc. 

Our motivation, in starting the following investigation of the SNR G18.95$-$1.1 was to solve, once and for all, the mystery of the energy source of the emission from the remnant, namely:  was a compact object powering the emission of the remnant? 
In view of the \asca\ data, we have reanalyzed the existing \rosat\ PSPC data, an analysis  we present in \S 2.  In \S 3 we describe the 
data extraction methods, and the spatial and spectral analysis of the \asca\ data. In \S 4 
we discuss the implications of our analysis and a summary of
our principal conclusions.

\section{Analysis of the \rosat\ PSPC data}
We extracted the 
G18.95$-$1.1 \rosat\ PSPC data from the public archive and 
processed them to minimize the contribution from 
the particle background \citep{sno94}. There is no \rosat\ HRI observation of 
this remnant.   
A point source, also mentioned in \citet{fur97}, 
is detected within the boundary of the SNR   
at 18$^{\rm h}$28$^{\rm m}$48$^{\rm s}$,
$-$13$^\circ$00$^\prime$55$^{\prime\prime}$ (J2000).  The source was not definitely identified by \citet{fur97} and it has no radio counterpart. 
This point source is positionally coincident (given the star's known proper motion and the position accuracy of the \rosat\  PSPC) with a star identified in 2MASS at a position of 
18$^{\rm h}$28$^{\rm m}$50.08$^{\rm s}$,  $-$13$^\circ$01$^\prime$20.3$^{\prime\prime}$ (J2000). We consider the likelihood that this is the counterpart to the X-ray source in more detail below. 

For our  \rosat\ PSPC spectral analysis of the entire remnant, we have extracted
data from a 17$^\prime$ radius circle (the extent of the 
remnant). In the first part of our 
analysis, we have modeled the thermal emission 
with collisional ionization equilibrium  (CIE) models, the 
so-called {\it mekal} model \citep{mew85, mew86, kaa92}  and the so-called {\it raym} model \citep{ray77} which are available in XSPEC v11.2, the 
X-ray spectral analysis package used throughout this analysis. 
In all the thermal models used, unless explicitly specified, 
we have kept the elemental abundances at
their solar  values as derived by \citet{and89}.
Absorption along the line of sight was taken into account 
with an equivalent column density of hydrogen, $N_{\rm H}$, using the 
cross-sections and abundances from the \citet{bal92} photo-electric 
absorption model. In agreement with existing results we found that a single component
thermal model can describe the data with a temperature around 0.3~keV and 
a hydrogen column density around 10$^{\rm 22}$~atoms~cm$^{\rm -2}$ 
for all the thermal models used in this analysis. 
 We examined non-equilibrium ionization (NEI) effects arising when the  
ions are not instantaneously ionized to their equilibrium 
configuration at the temperature of the
shock front \citep{gro82}. To incorporate
these effects into the model of SNR spectral emissivity, we use here 
the {\it pshock} model incorporated in XSPEC \citep{bor01}.  
This model incorporates the Fe-L shell atomic data computation from \citet{lie95}
and it is a 
first approximation to the physical phenomena which occurs at the shock. 
It does not include radiative and di-electronic recombinations, nor any coupling  between ions and electrons, but despite its shortcomings (present in most of the other models available), 
the model is useful to provide a good first approximation of the physical state of the plasma. The ionization state depends on the product of the electron density and the age, and we define the 
ionization timescale as  $\tau_i \equiv n_et$. The larger the value of the ionization timescale, the closer the system is to ionization equilibrium \citep{gro82}. 
When fitted with such a model, the data lead to  a similar, although slightly smaller, column density 
(7$\times$10$^{\rm 21}$~atoms~cm$^{\rm -2}$) but 
at a much higher temperature (3.3~keV) and an 
ionization timescale (n$_e$t = $\tau_i$ = 2.3$\times$10$^{\rm 10}$~cm$^{\rm -3}$~s)  far from the equilibrium  value (n$_e$t at equilibrium is larger than 10$^{\rm 12}$~cm$^{\rm -3}$~s). As found by \citet{fur97}, 
a non-thermal (power-law type) model does not fit the data
as well and the value of the spectral index found in this case 
($\alpha$ $>$ 7.5) is unphysically large.  
All fits (thermal and non-thermal) imply a large value of the 
absorption column density (close to 10$^{\rm 22}$~atoms~cm$^{\rm -2}$).\\

\section{\asca\  Analysis}
With a diameter of about 33$\myarcmin$,  G18.95$-$1.1 could not be entirely covered with the \asca\ SIS in the standard 2 CCDs mode (the GIS covers the remnant almost completely.) We chose to cover overlapping but different parts of the remnant with the two SIS detectors. The resulting field of view is a characteristic L-shaped image, the overlapping CCDs covering the region of the point source identified by \rosat. We used the standard processed data as provided by the \asca\ Guest Observer 
Facility\footnote{See the ``Guide for \asca\ data reduction'' at 
http://heasarc.gsfc.nasa.gov/docs/asca/ahp\_proc\_analysis.html
for more information on  the cuts applied to the data.}.
The event processing configuration was kept at its default value, with a  GIS time resolution of 0.125~s 
(the SIS is not used for timing analysis).
\subsection{Spatial}
We generated exposure-corrected, background-subtracted images of the
GIS and SIS data in soft (below 4.0~keV) and hard (above 4.0~keV) 
energy bands for the $\sim$  25~ks observation.\footnote{We used here a new 
\asca\ FTOOLS package called {\it fmosaic}.}
We used newly generated blank maps to estimate the 
background. These maps, available only for the GIS,  are more complete than the standard ``high latitude'' 
ones used in similar studies. They were first corrected for spurious point 
sources and then screened using the standard event screening 
criteria\footnote{For more details on the generation of these files, see 
http://heasarc.gsfc.nasa.gov/docs/asca/mkgisbgd/mkgisbgd.html.}.
We used the  standard ``blank maps'' for the SIS.  
Exposure maps (both for GIS and SIS) 
were generated using the {\it ascaeffmap} and {\it ascaexpo} 
scripts\footnote{Both scripts are part of the \asca\ FTOOLS package 
http://heasarc.gsfc.nasa.gov/docs/software/ftools/asca.html.}.
Events from regions of the merged exposure map with
less than 30\% of the maximum exposure were ignored.
Merged images of the source data, background, and exposure were smoothed with 
a Gaussian of standard deviation, $\sigma=45^{\prime\prime}$.  We
subtracted smoothed background maps from the data maps and divided by
the corresponding exposure map.  
Figure~2 shows the results of this procedure for 
both the SIS and the GIS detectors at energies below and above 4~keV. 
The GIS low-energy image, covering almost all of the remnant, 
shows two maxima of 
emission located respectively at 
18$^{\rm h}$29$^{\rm m}$15.6$^{\rm s}$, 
$-$12$^\circ$55$^\prime$22.5$^{\prime\prime}$ and 
18$^{\rm h}$28$^{\rm m}$49.9$^{\rm s}$, 
$-$13$^\circ$01$^\prime$7.44$^{\prime\prime}$ (2000). There is  evidence for emission above 4~keV in the GIS (a signal-to-noise ratio at the peak of about 4). This emission is centered at 
18$^{\rm h}$29$^{\rm m}$4.3$^{\rm s}$, 
$-$12$^\circ$52$^\prime$33.7$^{\prime\prime}$ and is confined to a small region (a little less than
3$^\prime$ in radius, comparable to the PSF of the instrument).  The SIS does not cover this part of the remnant.
We show in 
Figure~3 an overlay of the soft X-ray contours on the 30~GHz radio image with the location of the hard X-ray source marked by the small white circle. As seen in this picture, while the remnant's extent is slightly smaller in the X-ray band than in the radio, the central region shows obvious correlations in the emission morphology at both frequencies. 
The central radio bar overlaps almost entirely with the central low energy X-ray emission, and the maximum of the higher energy emission is located at the tip of that bar. 
\subsection{Spectral}
 We first studied the integrated emission from the remnant by extracting
the X-ray spectra from the GIS data from a circular region centered at 
18$^{\rm h}$29$^{\rm m}$23.72$^{\rm s}$,
$-$13$^\circ$00$^\prime$44.76$^{\prime\prime}$ (J2000) and using a 
radius of 11$^\prime$7.4$^{\prime\prime}$, chosen to encompass most of the emission from 
the SNR. This region covers almost the entire extent of the remnant in the  \rosat\ pointing, making 
a direct comparison between the two datasets possible.
We extracted the background from regions in the GIS field of view free of emission from the remnant.  Because the detector in the L-shaped configuration of the SIS does not cover the complete remnant, in the following analysis, we have used the SIS data exclusively in the study of the point source detected in the \rosat\ observation.  The measured count rates are  0.465$\pm$0.006~cnt~s$^{\rm -1}$ for 
the complete remnant (as measured with the GIS) and 
0.039$\pm$0.001~cnt~s$^{\rm -1}$ (0.042$\pm$0.001~cnt~s$^{\rm -1}$)  
for the point source in the GIS (SIS).  
Our spectral analysis is divided into several parts. We first study the complete remnant, then examine in detail the high-energy emission region detected above 4~keV in the GIS. 
We then analyze the data from the star and finally from the small part of the remnant where the radio emission is the strongest.  All detailed results are given in Tables 1 \& 2. 

We extracted the \asca\ GIS spectrum of the entire remnant (a total of 18600 events).
The gain  offset was allowed to  vary, and we measured a gain shift of -3.2\% and a range of variation between 
$-$2.6\% and $-$3.6\%. This gain adjustment  is consistent with the results from the 
  calibration data analysis done by the \asca\ GIS  team\footnote{see http://heasarc.gsfc.nasa.gov/docs/asca/ahp\_proc\_analysis.html for more information on calibration.}. 
We used a pure thermal (CIE ``mekal'') model to describe the emission 
and obtained a $\chi^2$ of 124.9 and a reduced $\chi^2$
($\chi^2_{\rm r}$) of 1.92.
We found a column density of 
9.8$\pm$0.5$\times$10$^{\rm 21}$~atoms~cm$^{\rm -2}$, 
completely consistent with the value derived from our
 \rosat\ analysis. The temperature  is $kT$~=~0.62$\pm$0.03~keV. 
 The unabsorbed flux in the [0.5-2.0]~keV range is between 1.9 and 
2.5$\times$10$^{\rm -10}$~ergs~cm$^{\rm -2}$~s$^{\rm -1}$, about 
three orders of magnitude higher than the contribution  
in the [4.0-10.0]~keV band. We then fitted the  \asca\ GIS and \rosat\ PSPC spectra 
simultaneously (the gain offset in the GIS was fixed to the value derived from 
the single GIS fit alone) as 
\rosat\ PSPC data provide a stronger constraint on the column density value. 
We found 8.4$\pm$0.3$\times$10$^{\rm 21}$~atoms~cm$^{\rm -2}$ associated 
with a similar temperature ($kT$~=~0.58$^{\rm +0.02}_{\rm -0.04}$~keV), 
very much consistent with the results found with the \asca\ GIS alone. As in the analysis of the \rosat\ PSPC, we examine nonequilibrium ionization effects using the {\it pshock} model \citep{bor01} described in the previous section.
As in the CIE model, all elements are kept to their  solar abundances as defined by \citet{and89} except when explicitly 
specified. The gain adjustment for the \asca\ GIS data is set to the same value as the one found
in the CIE analysis.  While we recognize that this is not formally correct (because NEI effects shift the line centroids relative to what is expected for a CIE model, precisely what our correction does), it is a  reasonable approximation (the shift between the CEI and NEI models is on the order of 1\% at Si-K, smaller than the gain shifts found in our analysis).  
A combined fit with the \asca\ GIS and the \rosat\ PSPC leads to a hydrogen column density 
9.4$^{\rm +0.03}_{\rm -0.10}\times$10$^{\rm 21}$~atoms~cm$^{\rm -2}$ 
consistent with the value found in the CIE fit. 
The associated temperature of 0.9$^{\rm +0.4}_{\rm -0.1}$ keV  is 
above the equilibrium value found in the previous analysis. 
This temperature from the NEI fit is  
associated with a low ionization timescale 
of (1.35$^{\rm +1.44}_{\rm -0.16}$)$\times$10$^{\rm 10}$~cm$^{\rm -3}$~s. The fit is significantly better than the one done using the equilibrium model (a drop in $\chi^2$ of 
more than 100 with the addition of one more degree of freedom) and this is the result that we will use in our analysis section. 
The results of all the fits are given in Table 1 and Figure~4 shows the result of the best {\it pshock} 
fit.  We have searched for possible deviations from solar abundance in the spectrum of the remnant. Any such variations could be the sign of either ejecta from the supernova explosion or anomalous abundances in the interstellar medium swept-up by the expanding blast wave. 
We find that departures from solar abundance for Mg, Si and S are not significant at the 90\% confidence level (CL). 
When allowed to vary, the  Mg abundance increases by about 50\% with a drop in $\chi^2$ of about 30, and the S abundance 
decreases to a very small value for a drop in $\chi^2$ of about 15.
The associated  column density, temperature, and ionization timescale   are either identical or compatible at the 90\% CL  with the values found in the previous analysis. We find a 
similar but somewhat less significant result (a much smaller drop in  
$\chi^2$) for the Si abundance.\\

We then analyzed the region of the high-energy emission. 
We first extracted,  for comparison, the spectrum from a region showing no sign of high-energy emission.  We chose a circle of about 2$^\prime$ radius and centered at 
18$^{\rm h}$29$^{\rm m}$28.935$^{\rm s}$, 
$-$13$^\circ$03$^\prime$52.47$^{\prime\prime}$ (J2000).  
The extracted spectrum from this region (a total of 690 events) is compatible with the results found for the whole remnant. We find that with the column density fixed to the value derived in the previous analysis, the temperature of this region is in the range 0.5 $-$1.6~keV and 
$n_et$ is essentially unconstrained. This poor precision is due to the small number of events in the spectrum.  We then extract the spectrum from a similar size region centered at 
18$^{\rm h}$29$^{\rm m}$4.3$^{\rm s}$, 
$-$12$^\circ$53$^\prime$33.7$^{\prime\prime}$ (J2000)  and used both the model derived from the study of the  "comparison" region (with an arbitrary  normalization value) and an additional model intended to characterize the excess emission at high-energy. 
The low-energy flux from the first region was used to normalize the underlying contribution of the rest of the remnant.  We find that, due mainly to the poor statistics on this spectra -- a total of 891 events, the excess X-ray high-energy emission cannot be characterized uniquely. 
Both thermal and non-thermal models can describe the data with a  $\chi^2_r$  around 2. 
In these two component models the column density is fixed to the value derived for the complete remnant (9.4$\times$10$^{\rm 21}$~atoms~cm$^{\rm -2}$). We find that a thermal model for the extra emission implies a relatively high temperature (a best fit at 2.1$^{\rm +7.5 }_{\rm -1.0 }$~keV for a Mekal model).  There are several tentative explanations for a small high-temperature region within a remnant. It could be a region of lower density shock-heated plasma, which assuming pressure equilibrium  would produce a higher measured temperature than elsewhere in the remnant. 
In principle, one could distinguish between different regions of emission but  \asca\ lacks the necessary spatial resolution.  A small region of ejecta material would be another  possible explanation  for a temperature variation on that scale. {\it ASCA}'s  shortcoming in this case, is its spectral resolution, too poor to identify abundance variation with the existing statistics. 
It could also be that the high-energy emission detected in the remnant  is powered by a hidden pulsar. In this case,  one would expect a non-thermal spectrum, the expected signature of the synchrotron emission from electrons accelerated in the nebula surrounding a pulsar. 
We find that for a non-thermal power law model of the emission,  the power law index for the best fit is  2.9$^{\rm +1.6}_{\rm -1.1}$ (at 90\% CL).  This is a bit larger than the canonical value but the large error bars associated with this result  preclude any definitive conclusion.  

We also studied the characteristics of the X-ray emission centered around the point source detected in the \rosat\ and \asca\ GIS and SIS data. 
We extracted the X-ray spectrum from a region centered at 
18$^{\rm h}$28$^{\rm m}$49.9$^{\rm s}$,  $-$13$^\circ$01$^\prime$7.44$^{\prime\prime}$ (J2000), the maximum of emission in the \rosat\ PSPC image.   
We find that the spectra (both \rosat\ PSPC and  \asca\ SIS) can be best described by a thermal model (CIE) at a temperature of 0.55$\pm$0.10 keV and a column density of 
9.2$^{\rm +1.3}_{\rm -1.1}\times$10$^{\rm 21}$~atoms~cm$^{\rm -2}$ 
consistent with the results of the general fit ($\chi^2_{\rm r}$= 1.18). 
As mentioned above, this source  is coincident with a star in the 2MASS and USNO-B1 catalogues. The observed infrared magnitudes from 2MASS are  J=9.854$\pm$0.027, H=9.444$\pm$0.036, and K=9.323$\pm$0.028. The star ( 5698-00714-1 in USNO-B1) has optical magnitudes of 
B=11.60$\pm$0.22 and a V=11.58$\pm$0.15.  We used the measured X-ray column density to deduce a reddening of A$_{\rm V}$=5.14$^{\rm +0.72}_{\rm -0.61}$  \citet{pre95}, which results in de-reddened magnitudes of B$_{\rm 0}$,  V$_{\rm 0}$, J$_{\rm 0}$, and K$_{\rm 0}$ of  4.85, 6.44, 8.27, and 8.60 respectively. 
The resulting (B-V)$_{\rm 0}$ color of 1.58 is inconsistent with any main sequence, giant, or supergiant stellar candidate, casting doubt on the identification of this star as the optical counterpart to the X-ray source.  A much lower value of A$_{\rm V}\sim$1 is required to make the
 colors compatible with an early-type star.  However our X-ray analysis clearly rejects a low column density; the lower limit allowed at 90\% confidence is 8.1$\times$10$^{\rm 21}$~atoms~cm$^{\rm -2}$ (A$_{\rm V}\sim$ 4.5).The definitive identification of the optical counterpart remains elusive. 

Finally, we also extracted a spectrum from the region of the brightest 
radio emission and the largest polarization and compared our results with those
obtained from the analysis of the complete remnant. The spectral extraction region was  about 
4$^\prime$ in radius and centered at 
18$^{\rm h}$29$^{\rm m}$17.16$^{\rm s}$, 
$-$12$^\circ$53$^\prime$52.20$^{\prime\prime}$ (J2000). The spectrum has fewer
than 900 events and we find that within 
the uncertainties, the X-ray emission from this region is
indistinguishable from the rest of the  remnant.  The values obtained for the hydrogen column density and the temperature are consistent
with those found for the rest of the remnant.  
Quantitative results and fluxes are tabulated in Table 2.\\

\section{ Discussion}   

As mentioned at the beginning 
of this paper, the distance to G18.95$-$1.1 is not well
known and we use the lower distance estimate (2~kpc) derived from 
 21~cm HI absorption data \citep{bra84,fur89}. 
At this distance, the supernova remnant defines an X-ray emitting volume of
 $V \simeq 
1.2\times10^{\rm 59}\,f\,D_{\rm 2}^{\rm 3}\theta_{\rm 17}^{\rm 3}$~cm$^{\rm -3}$, 
where $f$ is the volume filling
factor of the emitting gas within the SNR, and
 $\theta_{\rm 17}$ the angular radius in units of 17$\myarcmin$. 
Because of the large variations in the spectral results using both the 
standard CIE model and 
our {\it pshock} NEI analysis, all the following results were computed 
using error bars twice as large as that found in the analysis. This allows 
us to gauge more effectively the different 
evolution scenarios for G18.95$-$1.1 while
acknowledging the difficulty in reaching 
definitive conclusions with these data alone. We deduce a hydrogen number density 
$n_{\rm H}$ = (0.13 -- 0.24)$\,D_{\rm 2}^{\rm -1/2}\,\theta_{\rm 17}^{\rm -3/2}\,f^{\rm -1/2}$~cm$^{\rm -3}$.  The mass of  the X-ray emitting plasma  
$M_{X}$, is (17 -- 31) $\,D_{\rm 2}^{\rm 5/2}\,f^{\rm 1/2}\theta_{\rm 17}^{\rm 3/2}M_\odot$.
At a distance of 15~kpc this would imply a  mass of  3000 $M_\odot$, for a Taylor-Sedov value of 0.25 for the filling factor $f$ \citep{tay50,sed59} .  
With the preceding numerical values we estimate the supernova explosion energy $E$ to be
 (0.4 -- 0.9)$\times 10^{\rm 50}\,D_{\rm 2}^{\rm 5/2}\,\theta_{\rm 17}^{\rm 3/2}\,f^{\rm -1/2}$~ergs. 
For a nominal value of $f$ and at 2~kpc, E  is within 
(although on the low side) standard values for SN explosions. We note that this value is derived under  the hypothesis that electrons and ions are in temperature equilibrium (a simplification that may not be correct in this case) and that a larger explosion energy is allowed if the ions are substantially hotter than the electrons.  
An estimate of the age of the remnant can be made 
  using our spectral analysis results. We found $t$ between 4400 and 6100 yr, where 
the range reflects the uncertainties in the fit, but does not 
include any error on the distance (assumed at 2~kpc) -- 
the age varies linearly with the distance.  
The preshock ISM number density is 
 $n_{\rm 0} \sim \rho_0/{\rm m_{\rm H}}= (0.06\pm0.02)\,D_{\rm 2}^{-1/2}\,\theta_{\rm 17}^{\rm -3/2}\,f^{\rm -1/2}$~cm$^{\rm -3}$.  As noted in the introduction, 21~cm HI absorption data \citep{bra84, fur89} do not exclude a distance of 15~kpc  to G18.95$-$1.1.  If we compute the physical characteristics that this 
distance would imply for the remnant, we find an explosion energy larger
than  $E = 2\times 10^{\rm 52}$~ergs, one order of magnitude 
larger than the value derived 
from the current SN explosion models.  In addition, because of the large swept-up mass derived in that case, we argue that the results from our 
X-ray analysis actually rule out this distance.  Based on the radio data alone, it is not surprising that G18.95$-$1.1 was suspected to harbor a compact object.  As described in the spectral analysis summarized above (see \S 3), the X-ray emission does provide some clues to the real nature of the emission.
The X-ray emission from the remnant is clearly dominated by thermal emission and we argue that the remnant is in non-equilibrium ionization, a hypothesis  compatible with the age (between 4400 and 6100 yr) derived in the context of this model.   Our data reveal a clear region of high-energy emission (above 4~keV) located at the tip of the radio bar, although our spectral analysis does not lead to a definitive answer as to the origin of this emission. We  have examined two hypotheses. 
In the first scenario, this emission results from a high-energy nebula powered by a  hidden pulsar.  
Our X-ray analysis yields a non-thermal X-ray unabsorbed flux of  about 
7$\times$10$^{\rm 32}$ erg~s$^{\rm -1}$ between 0.2 and 4 keV  or 
about  0.7\% of the flux from the rest of the remnant. 
This, in turn, suggests a rotational energy loss \.{E} of about 7$\times$10$^{\rm 35}$ erg~s$^{\rm -1}$ \citep{sew88,bec97}.
By comparison,  this value is about twice that of the pulsar/SNR in W44 whose pulsar's luminosity is about 1\% that of the total remnant \citep{har96}. 
We  computed the radio luminosity associated with the region from which the high-energy emission is coming and we find about 2.5$\times$10$^{\rm 33}$ erg~s$^{\rm -1}$  at a distance of 2~kpc and using the spectral index from the bar $\alpha$=-0.22 given in \citet{fur97}. 
This value leads to a ratio $L_{r}$/$L_{x}$  about 3.6,  a bit larger than measured values  for other PWNe (that ratio varies from 3.4 for Vela to  3.2$\times$10$^{\rm -4}$ for MSH 15-52). 

We have presented the results of \rosat\ and \asca\ X-ray spectral and spatial 
studies of the SNR G18.95$-$1.1, a middle-aged SNR (about 5000 yr old) 
most probably in its Taylor-Sedov phase of evolution and at a distance of 2~kpc.
The remnant is best described by a nonequilibrium ionization model
with solar elemental abundances. 
There are several X-ray bright spots in the remnant that are generally consistent with
thermal emission. We find that we cannot make a definitive identification of the optical counterpart to the unresolved \rosat\ PSPC X-ray source. The nature of an unresolved hard \asca\ X-ray 
source located near the tip of the central radio bar remains enigmatic.
Future Chandra and XMM-Newton observations may shed more light on
this question.
\par 

\acknowledgments
Our research made use of data obtained from the High Energy
Astrophysics Science Archive Research Center Online Service, provided
by the NASA Goddard Space Flight Center. 
This publication makes use of data products from the Two Micron All Sky Survey, which is a joint project of the University of  Massachusetts and the Infrared Processing and Analysis Center, funded by the National Aeronautics and Space Administration and the National Science Foundation.  This research has also made use of the USNOFS Image and Catalogue Archive
   operated by the United States Naval Observatory, Flagstaff Station (http://www.nofs.navy.mil/data/fchpix/). IMH acknowledges support through NASA grants NAG5-8394. 
POS acknowledges support through NASA contract NAS8-39073 and grant NAG5-9281.
IMH would like to thank Steve Drake for his help in navigating the intricacies of star catalogues.

\clearpage

\begin{figure}
\plotone{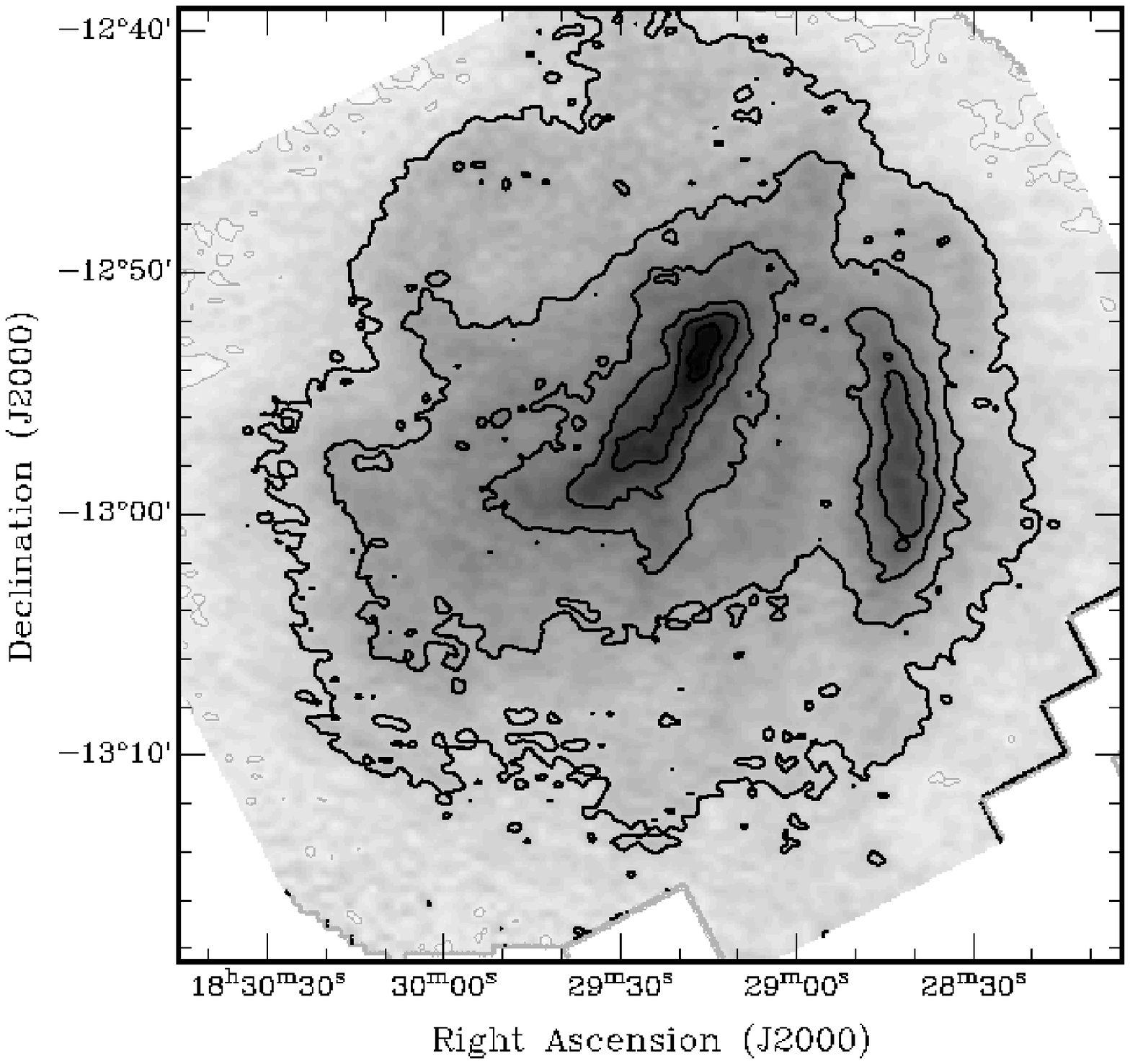}
\caption{\footnotesize The radio contour map of G18.95$-$1.1 at 10.55 GHz 
(image provided by Dr. F\"urst). The peaked intensity located at 
18$^{\rm h}$29$^{\rm m}$16.08$^{\rm s}$, 
$-$12$^\circ$53$^\prime$12.7$^{\prime\prime}$ is about 91 mJy. The remnant 
extent is about 17$^\prime$ in radius. \label{fig1}}
\end{figure}

\clearpage
\epsscale{.85}
\begin{figure}
\plotone{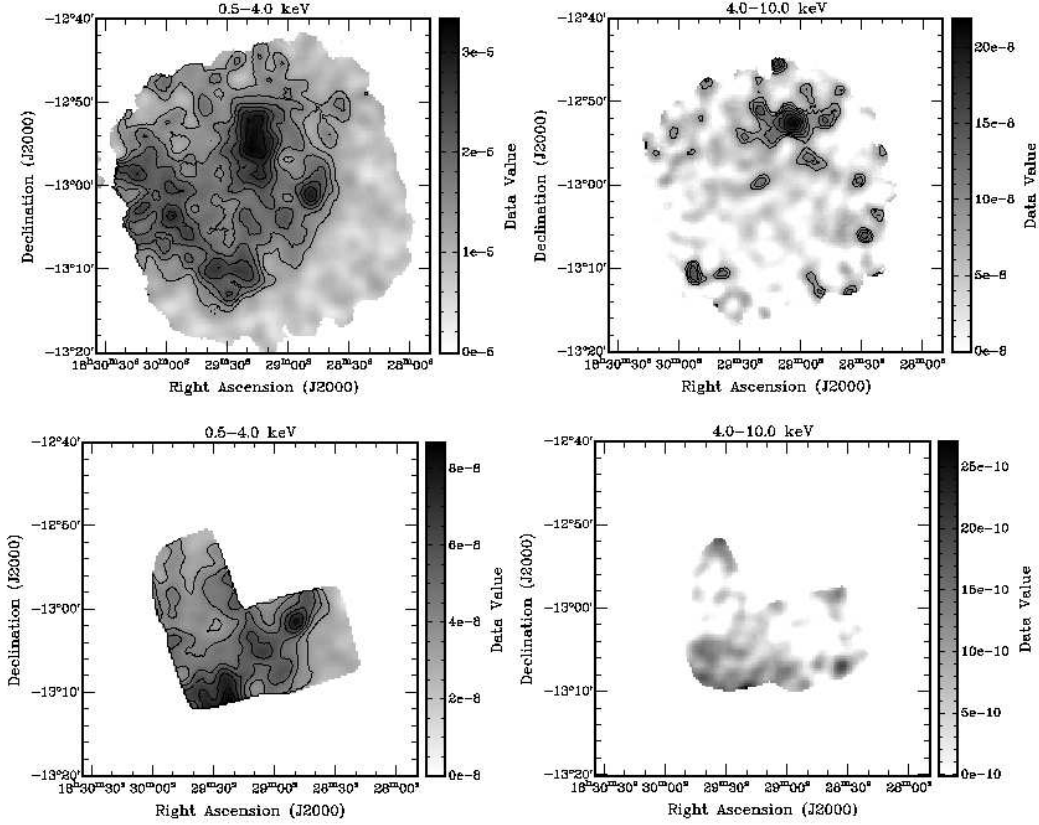}
\caption{\footnotesize  \asca\ X-ray images of the SNR G18.95$-$1.1 at low energy and high energy (0.5--4.0~keV and 4.0--10.0~keV) for the GIS (upper images) and the SIS
(lower images). We display only 6 linearly spaced contours, varying from 40\% to 90\% of the  peak
surface brightness in each map. The Peak/Background values are 4.2/0.7 (0.02/0.005) for the 
GIS at low (high energy). The similar value for the SIS are 7.5/1.6 at low 
energy (no high energy emission is detected in the SIS.) All values are quoted in units of 
$10^{-3}$cnt~s$^{-1}$~arcmin$^{-2}$.  
The bar on the left indicate the values associated with each maps in unit of cnt~s$^{\rm -1}$~pixel$^{\rm -1}$.  \label{fig3}}
\end{figure}

\clearpage
\begin{figure}
\epsscale{1.0}
\plotone{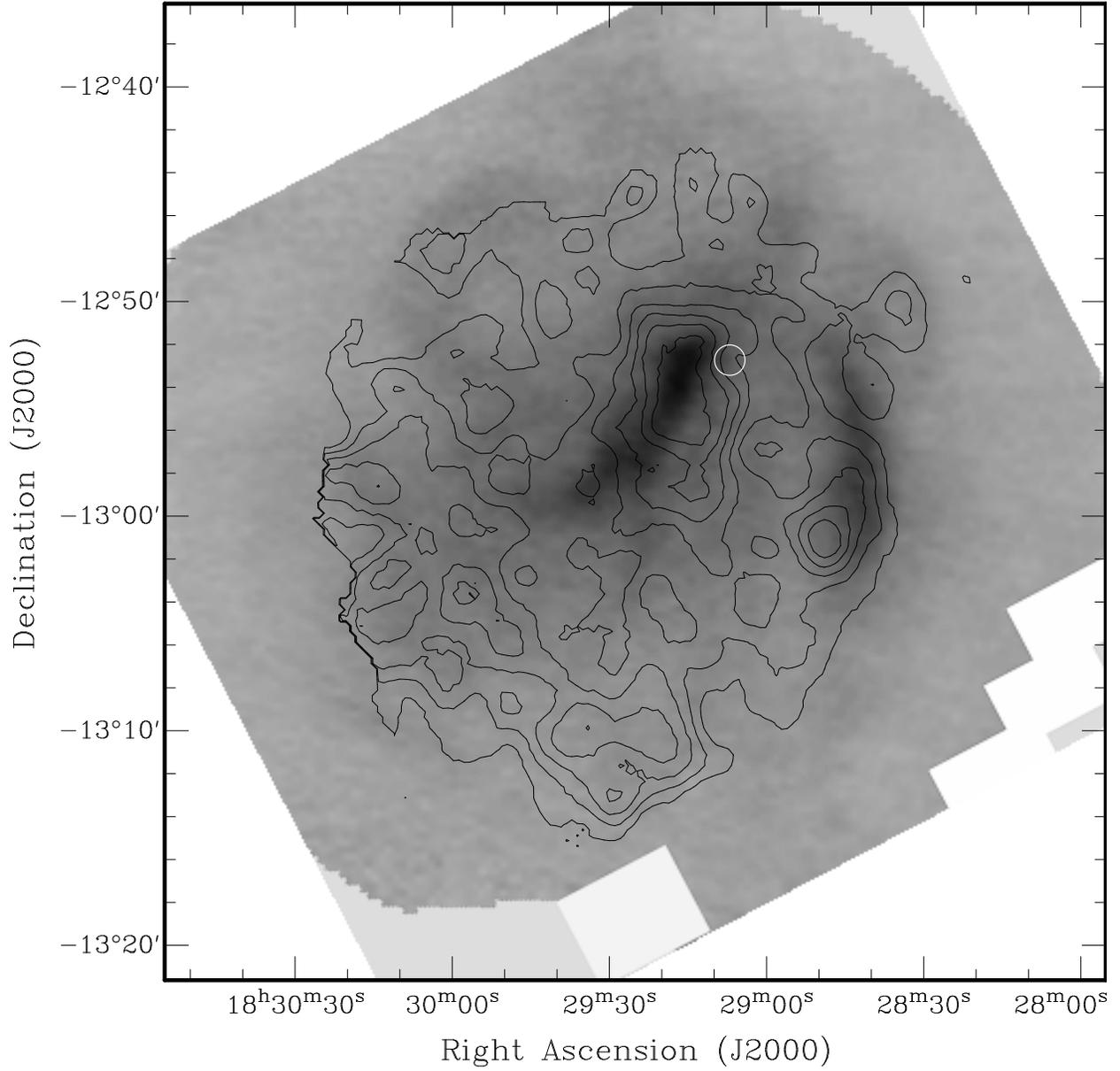}
\caption{\footnotesize  Radio image extracted 
from the Galactic Plane survey at 10.55~GHz (image provided by Dr. F\"urst), 
shown with contours derived from \asca\ GIS images at low  energy (0.5--4.0~keV). The position of the maximum of the high energy emission is marked by a small white circle. 
\label{fig4}}
\end{figure}

\clearpage
\begin{figure}
\plotone{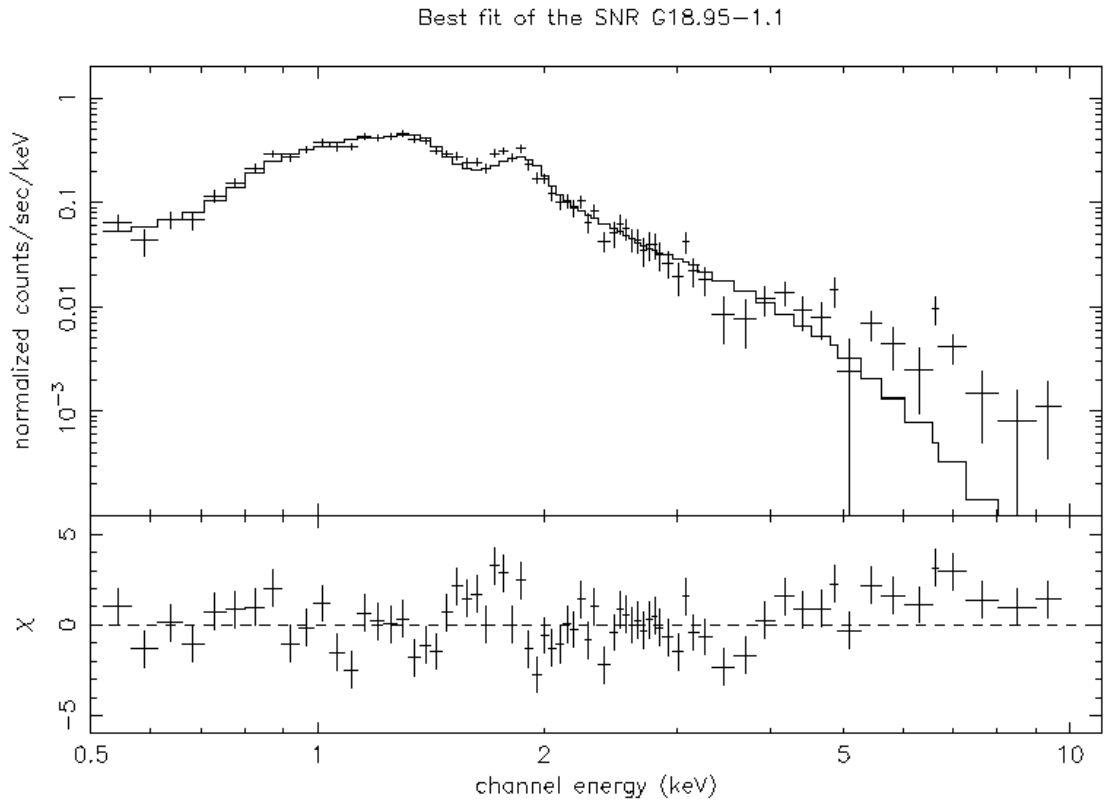}
\caption{ \footnotesize \asca\ GIS spectra of G18.95$-$1.1 extracted from a 11.1$\myarcmin\,$~~radius circular region. The solid curve in the top panel shows the best-fit using the {\it pshock} model (Borkowski et al. 2001).  The bottom panel shows the data/model residuals.\label{fig5}}
\end{figure}

\clearpage

\footnotesize
\centerline{\bf Table 1.}
\centerline{\bf Results from the Spectral Analysis}
\vspace{0.5cm}
\centerline{\begin{tabular}{lccc} \tableline\tableline \\[-10mm]
\multicolumn{1}{l}{Parameter} &
\multicolumn{3}{c}{Complete remnant (GIS and \rosat\ PSPC)} \\[1.5mm] 
\multicolumn{1}{l}{ }& 
\multicolumn{1}{c}{CIE thermal model$^{\rm a}$}& 
\multicolumn{1}{c}{NEI thermal model}& 
\multicolumn{1}{c}{NEI thermal model}\\[0.3mm]
\multicolumn{1}{l}{ }& 
\multicolumn{2}{c}{}& 
\multicolumn{1}{c}{variable Mg, Si, S abundances}\\[1.5mm] \tableline 
$N_{\rm H}$ (atoms~cm$^{-2}$)& 8.4$\pm$0.3$\times$10$^{\rm 21}$& 9.4$^{\rm +0.03}_{\rm -0.10}\times$10$^{\rm 21}$ & 8.3$^{\rm +0.4}_{\rm -0.5}\times$10$^{\rm 21}$\\[1.5mm]
$kT$ (keV) &   0.58$^{\rm +0.02}_{\rm -0.04}$ &   0.9$^{\rm +0.4}_{\rm -0.1}$ &   1.12$^{\rm +0.11}_{\rm -0.16}$\\[1.5mm]
log($n_et$) (cm$^{\rm -3}$~s) & N/A  & 10.13$^{\rm +0.31}_{\rm -0.06}$ &   10.15$^{\rm +0.11}_{\rm -0.09}$\\[1.5mm]
Normalization(cm$^{\rm -5}$)$^{\rm b}$ GIS &(7.7--9.3)$\times$10$^{\rm 12}$&(4.7--13.7)$\times$10$^{\rm 12}$ &  (4.2--8.4)$\times$10$^{\rm 12}$ \\[1.5mm] 
Normalization(cm$^{\rm -5}$) Rosat & (5.2--6.2)$\times$10$^{\rm 12}$ & (3.2--9.1)$\times$10$^{\rm 12}$  &(2.8--5.6)$\times$10$^{\rm 12}$\\[1.5mm] 
Flux (ergs~cm$^{\rm -2}$~s$^{\rm -1}$)~{\rm ([ 0.5 -- 2.0] keV)}  & (2.0--2.4)$\times 10^{\rm -10}$  &(0.5--1.5)$\times 10^{\rm -9}$ & (4.6--5.6)$\times 10^{\rm -10}$ \\[1.5mm]
Flux (ergs~cm$^{\rm -2}$~s$^{\rm -1}$)~{\rm ([ 2.0 -- 4.0] keV)}  & (3.3--4.5)$\times 10^{\rm -12}$   &(0.2--1.8)$\times 10^{\rm -11}$& (2.8--5.7)$\times 10^{\rm -12}$\\[1.5mm] 
Flux (ergs~cm$^{\rm -2}$~s$^{\rm -1}$)~{\rm ([ 4.0 -- 10.0] keV)}  & (0.67--1.4)$\times 10^{\rm -13}$ &(0.1--3.4)$\times 10^{\rm -12}$ & (0.32--1.10)$\times 10^{\rm -12}$ \\[1.5mm] 
Mg/Mg$_\odot$ & 1& 1& 1.5$\pm$0.2    \\[1.5mm] 
Si/Si$_\odot$ & 1& 1& 1.4$^{\rm +0.4}_{\rm -0.3}$    \\[1.5mm] 
S/S$_\odot$ & 1& 1& 0.4$^{\rm +0.6}_{\rm -0.4}$ \\[1.5mm] 
$\chi^{2}$/$\nu$ & 283.8/90 = 3.15  & 157.2/89 = 1.76& 116.8/86 = 1.36\\[1.5mm]\tableline
\multicolumn{2}{l} {$^{\rm a}$ Single-parameter 1~$\sigma$ errors}\\
\multicolumn{2}{l} {$^{\rm b}$ N=(${n_{\rm H}n_eV \over 4\pi D^2}$)}\\
\end{tabular}}

\vspace{2.5cm}
\footnotesize
\centerline{\bf Table 2.}
\vspace{0.5cm}
\centerline{\begin{tabular}{lccc} \tableline\tableline \\[-10mm]
\multicolumn{1}{l}{Parameter} &
\multicolumn{1}{c}{ Radio bright region (GIS)} &
\multicolumn{1}{c}{ Radio bright region (GIS)} &
\multicolumn{1}{c}{ Rest of the remnant (GIS)} \\[1.5mm] 
\multicolumn{1}{l}{ }& 
\multicolumn{1}{c}{CIE thermal model}& 
\multicolumn{1}{c}{NEI thermal model}&
\multicolumn{1}{c}{NEI thermal model}\\[1.5mm] \tableline 
$N_{\rm H}$ (atoms~cm$^{-2}$)& 1.6$\pm$0.4$\times$10$^{\rm 22}$& 0.6$^{\rm +0.4}_{\rm -0.3}\times$10$^{\rm 22}$ & 0.88$^{\rm +0.09}_{\rm -0.04}\times$10$^{\rm 22}$\\[1.5mm]
$kT$ (keV) &   0.30$^{\rm +0.12}_{\rm -0.06}$ &   0.50$^{\rm +0.3}_{\rm -0.1}$ &  0.80$^{\rm +0.4}_{\rm -0.1}$\\[1.5mm]
log($n_et$) (cm$^{\rm -3}$~s) & NA & $\leq$ 10.33 & $\geq$ 11.02 \\[1.5mm]
Normalization(cm$^{\rm -5}$)& (1.4--14.4)$\times$10$^{\rm 12}$ & (0.9--7.6)$\times$10$^{\rm 12}$ & (1.6--3.5)$\times$10$^{\rm 12}$ \\[1.5mm] 
Flux (ergs~cm$^{\rm -2}$~s$^{\rm -1}$)~{\rm ([ 0.5 -- 2.0] keV)}  &(0.22--3.3)$\times 10^{\rm -10}$ &(0.01--7.7)$\times 10^{\rm -10}$ & (4.5--7.3)$\times 10^{\rm -11}$\\[1.5mm] 
Flux (ergs~cm$^{\rm -2}$~s$^{\rm -1}$)~{\rm ([ 2.0 -- 4.0] keV)}  &(0.01--2.3)$\times 10^{-12}$ &(0.02--3.6)$\times 10^{\rm -12}$ & (1.0--7.2)$\times 10^{\rm -12}$\\[1.5mm] 
Flux (ergs~cm$^{\rm -2}$~s$^{\rm -1}$)~{\rm ([ 4.0 -- 10.0] keV)}  &(0.0006--1.8)$\times 10^{\rm -14}$  &(0.001--2.2)$\times 10^{\rm -13}$ &  (0.4--11.7)$\times 10^{\rm -13}$\\[1.5mm] 
$\chi^{2}$/$\nu$ & 33.22/27 = 1.23  & 23.00/26 = 0.88 & 265/135 = 1.95\\[1.5mm]\tableline
\end{tabular}}

\end{document}